\documentclass[aps,pre,twocolumn,floatfix,superscriptaddress,nofootinbib]{revtex4}
\usepackage{graphicx}
\usepackage{amsmath}

\begin{document} 

\title{Neural networks with excitatory and inhibitory components: direct and inverse problems by a mean-field approach}
\author{Matteo di Volo}
\email{mdivolo@iupui.edu}
\affiliation{Group for Neural Theory, Depart\'ement des Etudes Cognitives, Ecole Normale Sup\'erieure, Paris, France}
\affiliation{Centro Interdipartimentale per lo Studio delle Dinamiche Complesse, via Sansone, 1 - 50019 Sesto Fiorentino, Italy}
\affiliation{Indiana University Purdue University, Indianapolis IN, USA}
\author{Raffaella Burioni}
\email{raffaella.burioni@fis.unipr.it}
\affiliation{Dipartimento di Fisica e Scienza della Terra,  Universit\`a di
Parma, via G.P. Usberti, 7/A - 43124, Parma, Italy}
\affiliation{INFN, Gruppo Collegato di Parma, via G.P. Usberti, 7/A - 43124, Parma, Italy} 
\author{Mario Casartelli}
\email{mario.casartelli@fis.unipr.it}
\affiliation{Dipartimento di Fisica e Scienza della Terra,  Universit\`a di
Parma, via G.P. Usberti, 7/A - 43124, Parma, Italy}
\affiliation{INFN, Gruppo Collegato di Parma, via G.P. Usberti, 7/A - 43124, Parma, Italy} 
\author{Roberto Livi}
\email{livi@fi.infn.it}
\affiliation{Dipartimento di Fisica,  Universit\`a di Firenze, via Sansone, 1 - 50019 Sesto Fiorentino, Italy}
\affiliation{Istituto dei Sistemi Complessi, CNR, via Madonna del Piano 10 - 50019 Sesto Fiorentino, Italy}
\affiliation{INFN Sez. Firenze, via Sansone, 1 -50019 Sesto Fiorentino, Italy}
\affiliation{Centro Interdipartimentale per lo Studio delle Dinamiche
Complesse, via Sansone, 1 - 50019 Sesto Fiorentino, Italy}
\author{Alessandro Vezzani}
\email{alessandro.vezzani@fis.unipr.it}
\affiliation{ S3, CNR Istituto di Nanoscienze, Via Campi, 213A - 41125 Modena, Italy}
\affiliation{Dipartimento di Fisica e Scienza della Terra,  Universit\`a di Parma, via G.P. Usberti, 7/A - 43124, Parma, Italy}

\begin{abstract}
We study the dynamics of networks with inhibitory and excitatory leaky-integrate-and-fire neurons with short--term synaptic plasticity in the presence of  depressive and  facilitating mechanisms. The dynamics is analyzed by a Heterogeneous Mean-Field approximation, that 
allows to keep track of the effects of structural disorder in the network. We describe the complex behavior of different classes of excitatory and inhibitory components, that give rise to a rich dynamical phase--diagram as a function of the fraction of inhibitory neurons. By the same mean field approach,  we study and solve a global inverse problem:  reconstructing the degree probability distributions of the inhibitory and excitatory components and the fraction of inhibitory neurons from the knowledge of the average synaptic activity field. This approach unveils new perspectives in
the numerical study of neural network dynamics and in the possibility of using these models as testbed for the analysis of experimental data.
\end{abstract}

\maketitle

\section{Introduction}
Many of the brain activities emerge as the combined effect of excitatory and inhibitory components associated to synaptic plasticity
\cite{Royer,shadlen,sillito,wallace,tahv,brunel}. In mammalians, the fraction of inhibitory neurons is  close to 20-30\% \cite{noback}, and it seems plausible that this value has been 
determined by  evolutionary constraints, aiming at the  effectiveness of brain functions. Recently, an explanation has been
proposed referring to the possibility that such a rate between inhibitory and excitatory neurons could optimize the performances of
a neural network \cite{CHA}. Neurons in cortical area can exhibit quite complex scale-free structures, where inhibitory
neurons can play the role of hubs that control and moderate the action of the excitatory ones \cite{Boni}.
All of these considerations indicate that models of neural networks aiming at reproducing a great deal of 
brain functions should take into account the presence of both excitatory and inhibitory neurons, organized on a suitable network \cite{wilson,vanvree,brunel1,tsoinhib}. 

The large number of units and the typical high density of connections in many brain areas suggest that a mean field approach can be a proper mathematical tool for understanding the large scale dynamics of neural network models \cite{brunel,polmf,millman,mfcessac,mfbress}. Recently, we have applied a {\sl Heterogeneous Mean-Field} (HMF) strategy to deal with the dynamics of an excitatory neural network. This method retains the basic information on the network topology through the probability  distribution $P(\tilde k)$ of the in--degree density $\tilde k$ of synaptic connections attributed to each neuron, and it allows to build the dynamics of {\sl  classes} of neurons sharing the same in--degree density $\tilde k$, by a suitable discretization of the dynamical rule \cite{BCDLV}. The HMF approach is very effective in reproducing the main dynamical features of random dense networks of leaky-integrate-and-fire (LIF) excitatory neurons with synaptic short-term plasticity. 
In particular, the structure of quasi-synchronous events and the distinction between families of locked and unlocked neurons, with a rich and complex phenomenology  in synchronization related to the topological features of the network, are fully recovered \cite{BCDLV}.  


Interestingly, the HMF approach also  allows to
solve in a natural way a {\sl global inverse problem}. This consists in recovering the unknown 
degree probability distribution $P(\tilde k)$ from the knowledge of the average synaptic activity field. 
The method has been successfully applied to Gaussian and broad degree probability distributions of excitatory neurons, 
and it has been shown to be robust with respect to 
the introduction of noise and disorder \cite{lungo}. 

In this paper we show that the HMF strategy can be generalized to networks
of excitatory and inhibitory neurons, organized in a complex network topology, combining depressive and facilitating
mechanisms.  The main technical difficulty to overcome 
is that short-term synaptic plasticity obeys different dynamical rules for excitatory and
inhibitory neurons \cite{plast1,plast2}. Moreover, as discussed in Sec. \ref{uno},  one has to distinguish between the
dynamics of postsynaptic  excitatory and inhibitory neurons and, for both of these subclasses, between
signals coming from presynaptic excitatory and inhibitory  neurons. 

A comparison between the original network dynamics and the corresponding HMF dynamics is reported in Sec. \ref{due}. There  we show that, even on a random network with both excitatory and inhibitory component,  the HMF  approach reproduces the main features of the different complex dynamical regimes.

In Sec. \ref{tre} we discuss the features of the global synaptic activity fields emerging from 
the HMF dynamics for different values of the inhibitory fraction. In particular, we show that the system can display a quasiperiodic behavior characterized by locked and unlocked neurons, or an asynchronous regime where all neurons have  different oscillation frequencies. Moreover, for a specific value of the inhibitory fraction, the system features an {\it optimal} synchronization regime, where all neurons display the same interspike interval. 

In Sec. \ref{quattro} we derive an analytic relation between excitatory and inhibitory global synaptic activities, and we use this result to show, in Sec. \ref{cinque},  that the global inverse problem can be solved also for neural networks containing both excitatory and inhibitory components, with Gaussian and scale--free degree density distributions. In particular, we are able to reconstruct, from the average synaptic activity, the degree density distributions and the inhibitory fractions on a network with 10\% inhibitory neurons and  two gaussian distributions $P(\tilde k)$ for both inhibitory and excitatory components. The same holds in the case of a network with 30\%  inhibitory neurons generated by two $P(\tilde k)$  scale--free distributions, with a larger average value of $\tilde k$ for the inhibitory components. 

Conclusions and perspectives of our research are finally presented in Sec. \ref{sei}.

\section{LIF excitatory and inhibitory neurons with synaptic plasticity }
\label{uno}
We consider a network of $N$ neurons, either excitatory or inhibitory. Calling $v_i(t)$ the membrane potential of neuron $i$, its dynamics is ruled by the LIF model, i.e. 
\begin{equation}
\label{eq0}
\dot v_i(t)=a-v_i(t) +I^{syn}_i(t)~,
\end{equation} 
where $a$ is the common leakage current, and $I^{syn}_i(t) $ is the synaptic current coming from the connections with other neurons. All  variables can be rescaled  to work with adimensional units (see \cite{BCDLV, lungo}). 
For instance, time is rescaled to the membrane time constant $\tau_m=30$ms, and the spiking threshold $v_{th}$ of $v$ is set to 1, while its reset value is $v_r=0$. Whenever $v_i$ reaches $v_{th}$,  the neuron $i$ emits a spike and is reset to  $v_r$.  In our simulations we set $a=1.3$, so that neurons are in a spiking regime, i.e. even in absence of synaptic stimuli they fire periodically with a  period $T_0=\mathrm{ln}(a/(a-1))$. For the coupling dynamics we use the Tsodyks, Uziel and Markram (TUM) model, a description of short term synaptic plasticity that has been successfully tested in experimental setups \cite{plast1,plast2}. According to \cite{tsodyksnet}, the dynamics of the synapse between postsynaptic (i.e., receiving) neuron $i$ and presynaptic (i.e., transmitting) neuron $j$  is described in terms of  
the fraction of its {\sl active}, $y_{ij}(t)$, {\sl available}, $x_{ij}(t)$, and {\sl inactive}, $z_{ij}(t)$, resources. These 
quantities are  assumed to evolve according to the  following set of
coupled  differential equations:
\begin{align}
\label{eq1}
& \dot y_{ij}(t) = -\frac{y_{ij}(t)}{\tau_{\mathrm{in}}} +u_{ij}(t)x_{ij}(t)S_{j}(t)\\
\label{contz}
& \dot x_{ij}(t) = \frac{z_{ij}(t)}{\tau^{i}_{\mathrm{r}}}  -  u_{ij}(t)x_{ij}(t)S_{j}(t) \\
\label{eq3}
& x_{ij}(t)+y_{ij}(t)+z_{ij}(t)=1,
\end{align}
where the last equation is a conservation rule and $S_j(t)=\sum\delta(t-t_j(n))$ is the spike train of the presynaptic neuron $j$ emitting its $n$--th pulse at time $t_j(n)$. Whenever neuron $j$ emits a spike, it activates a fraction $u_{ij}$ of the available resources $x_{ij}$. In between two consecutive spikes, the fraction of active resources $y_{ij}$ decreases in time with a time constant $\tau_{\mathrm{in}}$, and the fraction of available resources recovers in a time $\tau^{i}_{\mathrm{r}}$ the fraction of inactive resources $z_{ij}$. If the postsynaptic neuron $i$ is inhibitory, the recovery time is much smaller. In particular, the typical phenomenological values are $\tau_{\mathrm{in}}=0.2$, while $\tau^{i}_{\mathrm{r}}=3.4$, if $i$ is inhibitory, and $\tau^{i}_{\mathrm{r}}=26.6$, if $i$ is excitatory. Moreover,  if the index $i$ corresponds
to an excitatory neuron,  $u_{ij}(t)$ is assumed to be constant, namely $u_{ij} = U = 0.5$, otherwise
\begin{equation}
\label{ufinite}
\dot u_{ij}(t) =
 - \frac{u_{ij}(t)}{\tau_{\mathrm{f}}} +U_f(1-u_{ij}(t))S_{j}(t) 
\end{equation}
where $\tau_{\mathrm{f}}=33.25$ is the facilitation time scale and $U_f=0.08$ is a phenomenological parameter \cite{tsodyksnet,volman} . 

The TUM model equations combine depressive and facilitating mechanism of plasticity \cite{tsodyksnet} . If the postsynaptic neuron is excitatory, the mechanism is purely depressive as a high frequency spiking of presynaptic neurons delays the available synaptic resources. If the postsynaptic neuron is inhibitory, the dynamics of $u_{ij}$ describes a facilitating mechanism, reinforcing the synapse when presynaptic neuron $j$ has a high electric activity. Equations 
(\ref{eq1})-(\ref{ufinite}) participate in the neural network dynamics by specifying  in Eq.(\ref{eq0}) the form of the synaptic current
received by neuron $i$:
\begin{equation}
\label{eqsyn}
I^{syn}_i(t)  =\frac{g}{N} \sum_{j \ne i} \epsilon_{ij} y_{ij}(t) 
\end{equation} 
where $g$ is the coupling parameter and the index $j$ labels the presynaptic neurons of neuron $i$.
The matrix elements  $\epsilon_{ij}$ can take the values 0, 1 and -1 if, respectively,  presynaptic neuron $j$ is disconnected,
excitatory or inhibitory with respect to postsynaptic neuron $i$.  
In our first approach, we consider random uncorrelated dense networks, i.e. networks where any correlation among 
different degrees is absent, and the degree is proportional to $N$ (this explains the  normalization factor 
$1/N$ in Eq.(\ref{eqsyn})~).

By defining $\tilde k=k/N$ the rescaled in--degree, where $k\in[0,N-1]$ is the number of in-connections of a given neuron, we can associate to the uncorrelated network its in-degree distribution $P(\tilde k)$. In general, inhibitory and excitatory neurons may have different in--degree distributions, $P_{I}(\tilde k)$ and $P_{E}(\tilde k)$. In particular, $P_{I}(\tilde k)$ and $P_E(\tilde k)$ are the probabilities that an inhibitory and an excitatory neuron receives $\tilde k\/N$ inputs from other neurons. In this setup, we fix the in-degree distributions of inhibitory and excitatory neurons, assuming that typically they have randomly distributed outputs.

\section{Heterogeneous mean field for a network of excitatory and inhibitory neurons}
\label{due}

The HMF approach (see \cite{BCDLV, lungo})  amounts to perform a thermodynamic limit, while keeping the 
neuron in--degree
density $\tilde k$ fixed. Dynamics (\ref{eq0}) is replaced by an evolution rule for {\sl classes} of neurons
labeled by their $\tilde k$
\begin{equation}
\label{eqHMF}
\dot v_{\tilde k}(t)=a-v_{\tilde k}(t) + g {\tilde k} Y(t)
\end{equation}
where $Y(t)$ is the average synaptic activity field.  In \cite{BCDLV, lungo} this method was applied to
dense uncorrelated  networks of excitatory LIF neurons, and it revealed a very good approximation
of dynamics ({\ref{eq0}) for any large finite network. For instance, the HMF approach is
effective also for sparse uncorrelated networks, provided they exhibit a sufficiently large
average in--degree \cite{lungo}.

Here we describe how the HMF approach can be extended to networks made  of inhibitory and 
excitatory neurons.  In this case,  the dynamics of each neuron depends on the
number of  its inhibitory and excitatory presynaptic neurons. This information is stored in 
the  adjacency matrix $\epsilon_{ij}$, that encodes the network topology.  Following the HMF
strategy, we have to split dynamics (\ref{eqHMF}) into two equations for classes of excitatory ($E$)
and inhibitory ($I$) postsynaptic neurons with in--degree density $\tilde k$:
\begin{align}
\label{vk1}
& \dot v^E_{\tilde k}(t)= a -v^E_{\tilde k}(t) +g\tilde k(-f_IY_{EI}(t)+f_EY_{EE}(t))\\
\label{vk2}
& \dot v^I_{\tilde k}(t)= a -v^I_{\tilde k}(t) +g\tilde k(-f_IY_{II}(t)+f_EY_{IE}(t)) \quad .
\end{align}
The last expressions on the r.h.s. of these equations correspond to the 
the average synaptic activity fields  received  by postsynaptic $E$ and $I$ neurons 
with label $\tilde k$ from their presynaptic $E$ and $I$ neurons.
The fractions of inhibitory and excitatory neurons are
denoted by $f_I$ and $f_E=1-f_I$, respectively.

These equations determine the spike trains $S^{I}_{\tilde k}(t)$  and $S^{E}_{\tilde k}(t)$ of inhibitory and excitatory 
classes of neurons, with in--degree density $\tilde k$.  On their side, these quantities enter the set of Eq.s
(\ref{eq1})-(\ref{eq3}), that split into four sets, identified by the upperscript  $(\dagger,*)$, where both symbols
can be either $I$ (inhibitory) and $E$ (excitatory), with $"*"$ corresponding to presynaptic neurons with
label $\tilde k$. In formulae 
\begin{align}
\label{yk}
& \dot y^{(\dagger, *)}_{\tilde k}(t) = -\frac{y^{(\dagger, *)}_{\tilde k}(t)}{\tau_{\mathrm{in}}} +u^{(\dagger, *)}_{\tilde k}
(t) x^{(\dagger, *)}_{\tilde k}(t) S^{*}_{\tilde k}(t)\\
\label{contz}
& \dot x^{(\dagger, *)}_{\tilde k}(t) = \frac{z^{(\dagger, *)}_{\tilde k}(t)}{\tau^{\dagger}_{\mathrm{r}}}  -  u^{(\dagger, *)}_{\tilde k}(t) x^{(\dagger, *)}_{\tilde k}(t) S^{*}_{\tilde k}(t) \\
\label{constr}
& x^{(\dagger, *)}_{\tilde k}(t)+y^{(\dagger, *)}_{\tilde k}(t)+z^{(\dagger, *)}_{\tilde k}(t) =1 ,
\end{align}
where $u^{(\dagger, *)}_{\tilde k}=U$ if $\dagger = E $, otherwise 
\begin{align}
\label{uk}
&\dot u^{(\dagger, *)}_{\tilde k}(t) =-\frac{u^{(\dagger, *)}_{\tilde k}(t)}{\tau_{\mathrm{f}}} +U_f\Big(1-u^{(\dagger, *)}_{\tilde k}(t)\Big)S^*_{\tilde k}(t).
\end{align}
Notice that the value of parameter  $\tau^{\dagger}_{\mathrm{r}}$ depends  on the  type of the postsynaptic neuron.

These equations can be closed by the consistency relations defining the  average fields that appear in
Eq.s (\ref{vk1}) and (\ref{vk2}):
\begin{equation}
\label{ygrandestar}
Y_{\dagger, *}=\int^1_0 P_{*}(\tilde k) y^{(\dagger, *)}_{\tilde k} d\tilde k \quad .
\end{equation}
For what follows, it is convenient also to define the global average fields, $ Y_I$ and $Y_E$, received 
by inhibitory and excitatory neurons:
\begin{align}
\label{globI}
& Y_I=-f_IY_{II}(t)+f_EY_{IE}~, \\
\label{globE}
& Y_E=-f_IY_{EI}(t)+f_EY_{EE} ~.
\end{align}

\section{Dynamical effects of inhibitory neurons}
\label{tre}
In a series of  papers \cite{DLLPT,BCDLV,lungo} we have analyzed in detail the dynamics of
random, uncorrelated, dense networks of excitatory LIF neurons and successfully compared it 
with the corresponding HMF dynamics. 
Analogously, in this section, we provide a short summary 
of some basic dynamical regimes of the HMF dynamics with inhibition 
presented in Section \ref{due}. 
In order to appreciate the reliability of this 
approach,  we preliminarily give a comparison between  the HMF results and direct numerical
simulations, performed on a large finite network made of  $N=5000$ neurons.

\begin{figure}
\centering
\includegraphics[width=7.5 cm]{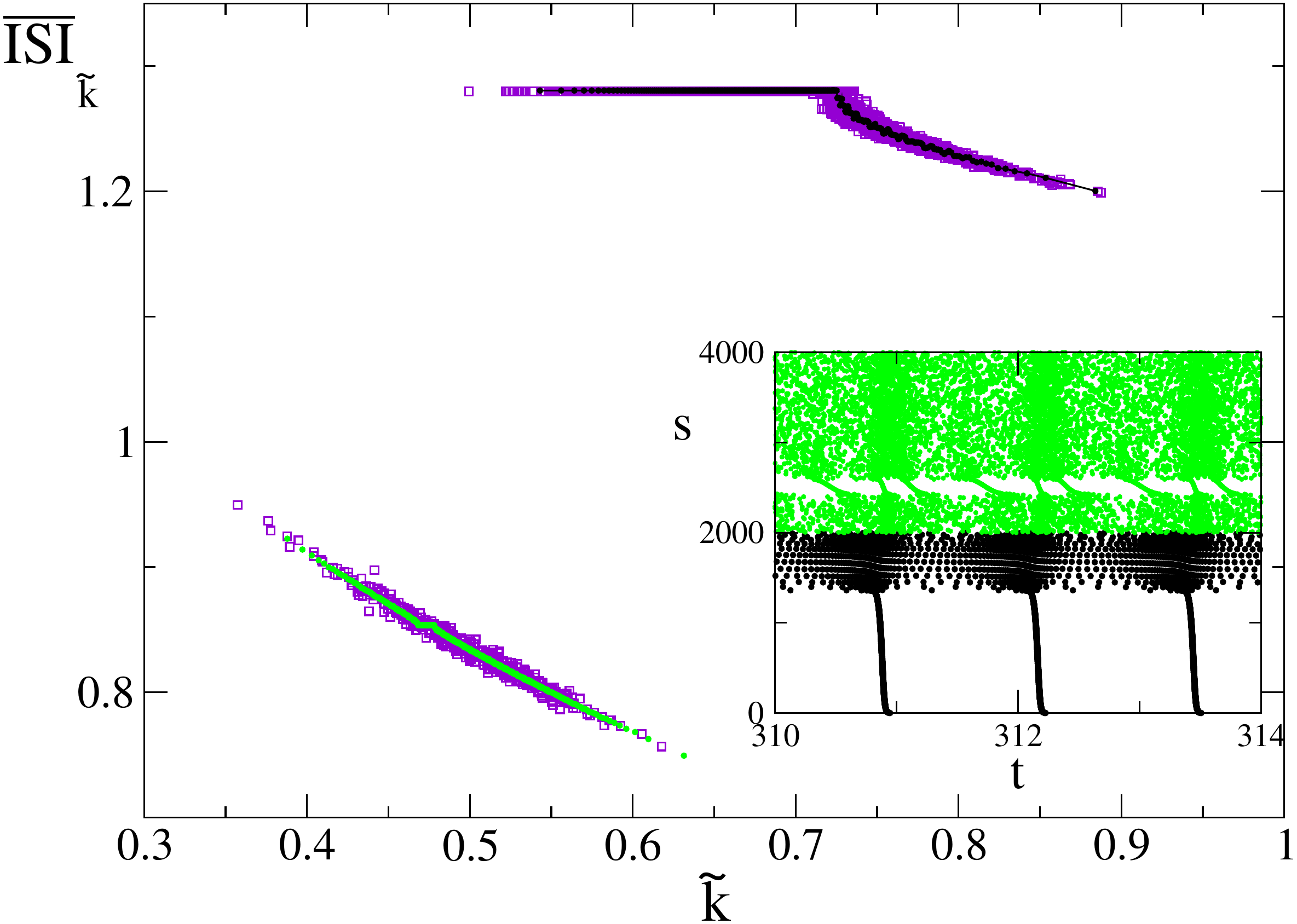}
\caption{(Color online) Average inter--spike time interval ($\overline{ISI}$) for a network of  $N=5000$ neurons (violet squares) with $f_I=0.1$
as a function of the in-degree density $\tilde k$.
The distributions $P_I(\tilde k)$ and $P_E(\tilde k)$ are both Gaussian with  
$\langle \tilde k_I\rangle=0.5$,  $\langle \tilde k_E\rangle=0.7$, $\tilde \sigma_I=0.04$ and $ \tilde \sigma_E=0.056$,
respectively. 
The black (relative to excitatory neurons) and green (relative to inhibitory neurons) dots have been obtained by the corresponding HMF dynamics. Notice the plateau region
typical of the population of excitatory neurons, that is almost absent for the population of inhibitory neurons.
The inset shows the raster plot of the HMF dynamics, where the neuron index $s$ is ordered according
to the in--degree density $\tilde k$ (see text) : excitatory neurons correspond to black dots ( $0 < s \leq  2000$) while inhibitory neurons to green dots. } 
\label{isiconfr}
\end{figure}

In Fig.  \ref{isiconfr} we plot the average 
inter--spike time interval ($\overline{ ISI}_{\tilde k}$),  of each neuron as a function of its in--degree density $\tilde k$. 
Data  have been obtained  for Gaussian probability distributions 
$P_I(\tilde k)$ and  $P_E(\tilde k)$ of $\tilde k$, 
with $\langle \tilde k_I\rangle=0.5$,  
$\langle \tilde k_E\rangle=0.7$, and standard deviations $\tilde \sigma_I=0.04$ and $ \tilde \sigma_E=0.056$.
Moreover, here and in the following Figures, we have assumed the  phenomenological values 
of the parameters $\tau_{\mathrm{f}}=33.25$  and $g=30$ (see \cite{tsodyksnet,volman}). 
The matching between the HMF dynamics and direct simulations is remarkable. 
As already observed in fully excitatory networks \cite{BCDLV, lungo}, excitatory neurons split into two families, 
namely periodic (locked)  and aperiodic (unlocked) neurons,  respectively observed for 
$\tilde k<\langle \tilde k_E \rangle$ and  for $\tilde k>\langle \tilde k_E \rangle$. 
Inhibitory neurons cover an approximately uniform  range of  higher frequencies. In the inset we report the raster plot 
(i.e., index of the firing neuron $s$ vs. its firing time $t$)
of the HMF dynamics to point out  the microscopic organization of  neurons. Excitatory neurons
correspond to $0\leq s \leq 2000$. In practice, the HMF dynamics has been obtained by sampling
$\tilde k$ with 2000 values for both groups of neurons, and $s$  has been ordered according to 
decreasing  values of $\tilde k$. In fact, the quasi--synchronous bursts observed for $0 \leq s < 1300$
are produced by the {\sl locked} excitatory  neurons in the plateau region. The {\sl unlocked} ones
($1300 \leq s < 2000$) exhibit quite irregular firing behavior, as well as most of the inhibitory neurons, 
that fire more frequently thanks to the facilitation mechanism typical of their synaptic activity. 
Only a small fraction of inhibitory neurons,
whose $\tilde k$ values overlap with those of excitatory neurons in the plateau, produce quasi--synchronous
bursts, but  with a different pace and regularity with respect to the excitatory ones.

 We are facing a sort of
of higher order locking effect of topological origin, induced by excitatory neurons in the plateau over equally coupled
inhibitory neurons. 
 Despite the complexity of the raster plot detailing this regime, the average activity fields $Y_E(t)$ and 
$Y_I(T)$ exhibit  periodic oscillations, that characterize a dynamical phase with a high level of synchrony, mainly
driven by  locked excitatory neurons  (e.g., see the inset in the upper panel  of Fig. \ref{isiconfr1} ) . 

\begin{figure}
\centering
\includegraphics[width=9.1 cm]{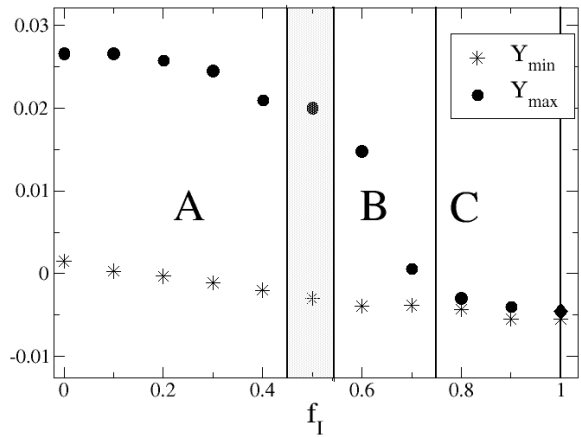}
\caption{ (Color online) Maximum (dots) and minimum (stars) values of the global field $Y_E(t)$ as a function of the fraction of inhibitory neurons $f_I$ obtained from the HMF dynamics. The distributions 
$P_{E/I}(\tilde k)$ and the HMF sampling adopted  are the same of Fig. \ref{isiconfr}.}
\label{sigma}
\end{figure}

\begin{figure}
\centering
\includegraphics[width=7.5 cm]{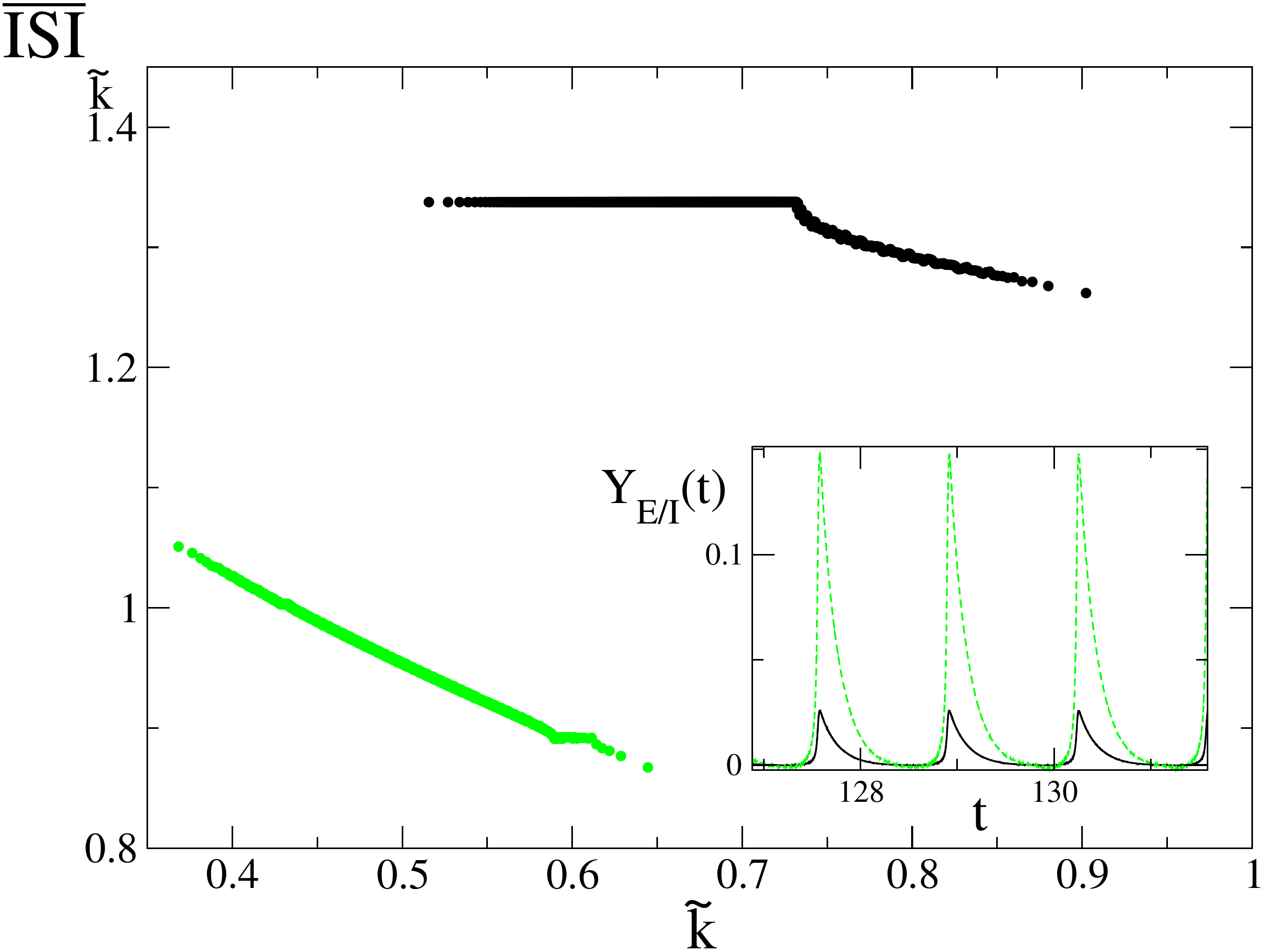}
\includegraphics[width=7.5 cm]{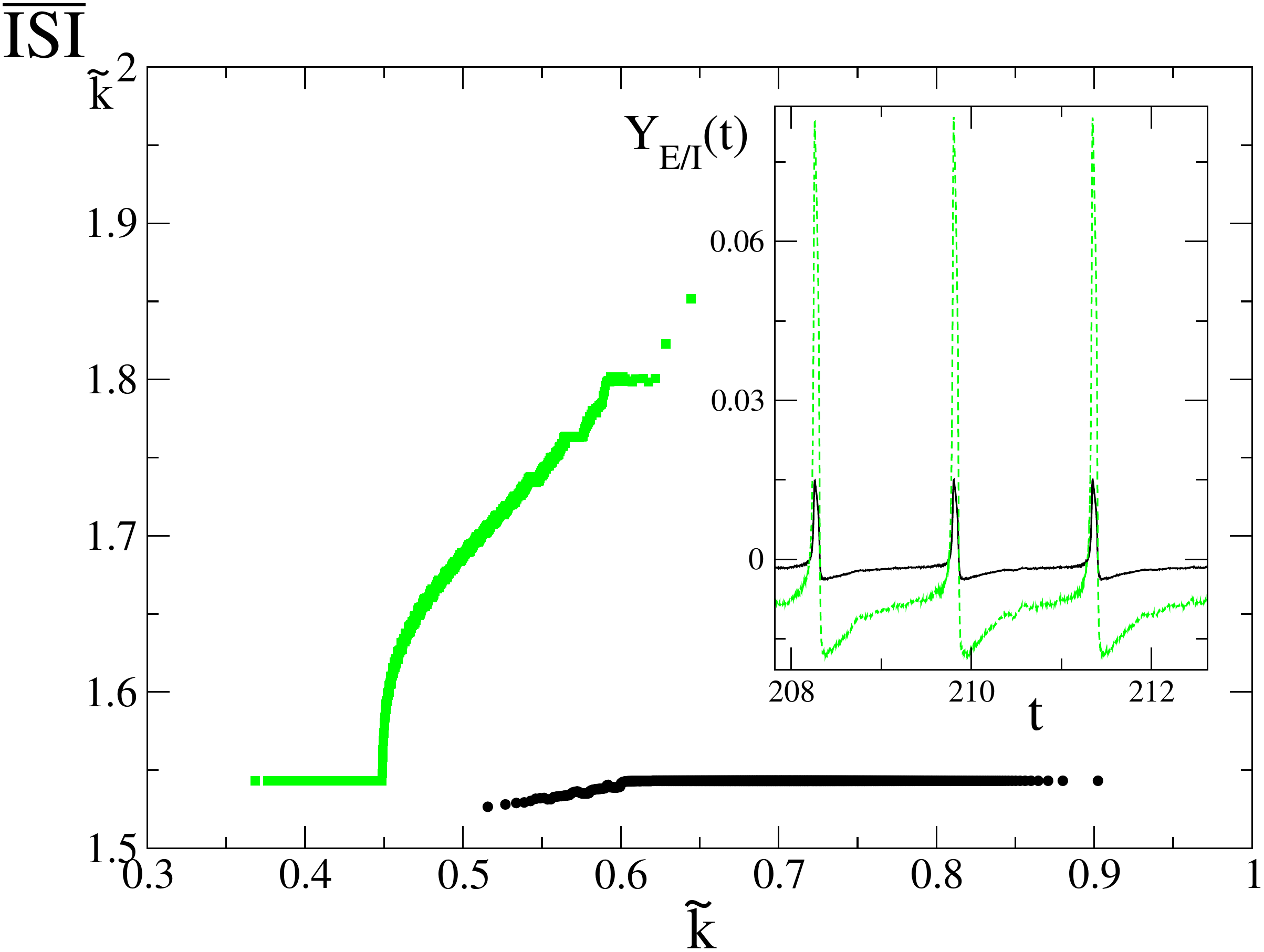}
\includegraphics[width=7.5 cm]{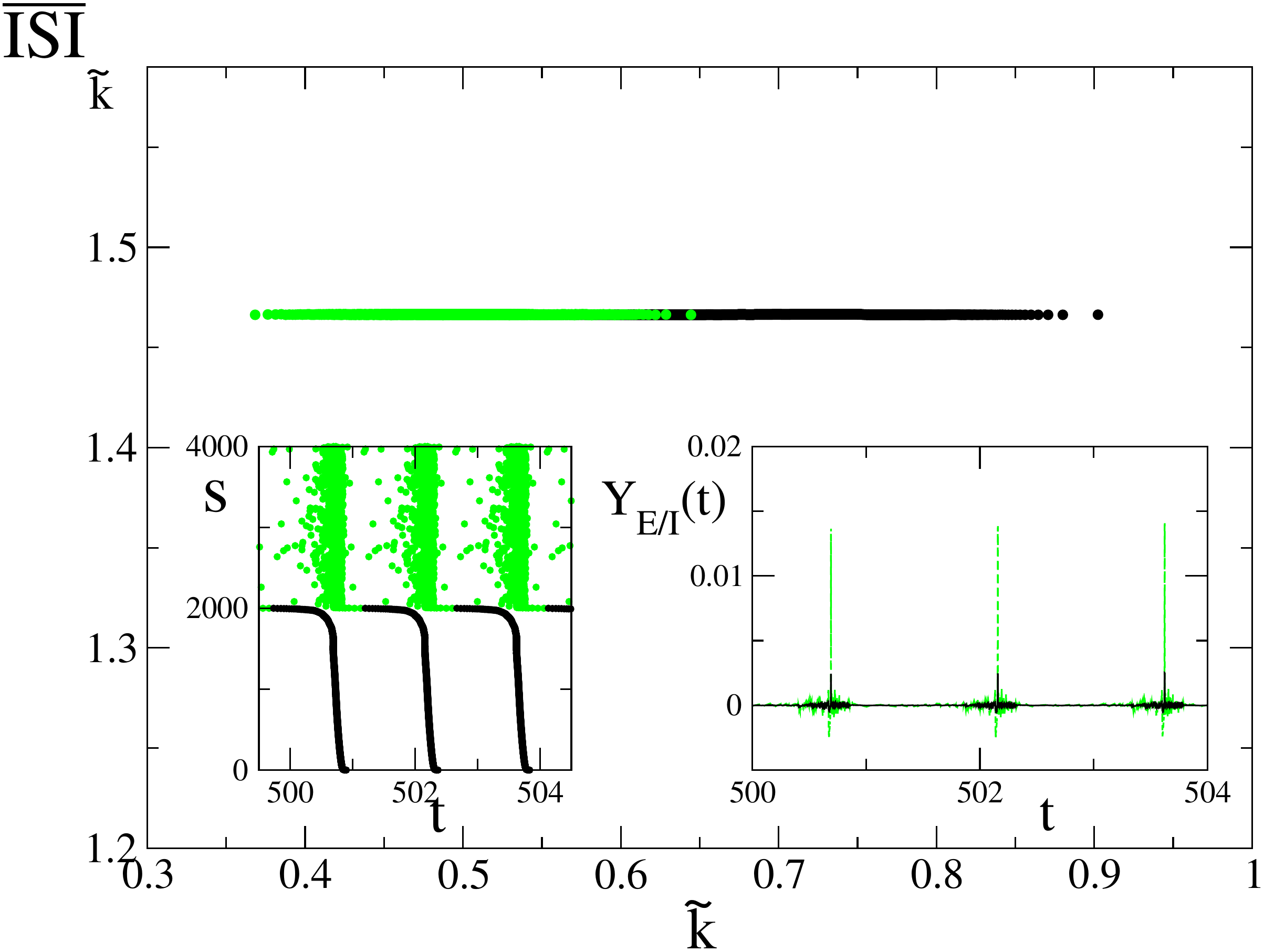}
\caption{(Color online) Dynamical regimes for different values of the fraction of inhibitory neurons $f_I$
for the same model of Fig.\ref{isiconfr}.
The three panels show the dependence of the average $\overline{ISI}_{\tilde k}$  on $\tilde k$ for $f_I = 0.2$
(upper panel), $f_I=0.6$ (middle  panel) and $f_I=0.5$ (lower  panel). Green dots stand for inhibitory neurons while black dots for excitatory neurons. The insets show the 
average activity fields $Y_E(t)$ (black lines) and $Y_I(t)$ (green dashed).
On the left of the lower panel there is an additional inset containing the raster plot of firing events. As in the inset of Fig. \ref{isiconfr}, black dots are relative to excitatory neurons and green dots to inhibitory neurons.}
\label{isiconfr1}
\end{figure}

In what follows,  we report how the dynamical regime of the HMF dynamics of the model described
in Fig. \ref{isiconfr} changes as a function of the fraction $f_I$ of inhibitory neurons. 
As pointed out in \cite{olmi}, an effective order parameter for 
exploring the HMF phase-diagram is given by the extremal values of $Y_E(t)$.  In fact, for increasing
values of $f_I$, the amplitude of $Y_E(t)$ decreases  and the  synchrony mechanisms 
inside  the network  are significantly modified.
In Fig. \ref{sigma} we plot the maximum (dots) and minimum (stars) values of  $Y_E(t)$ 
as a function of $f_I$. One can distinguish three main regimes. For $0 < f_I < 0.45$ (regime A), the network 
dynamics is driven by locked excitatory neurons. The upper panel of Fig. \ref{isiconfr1}
shows the average ISI as a function of $\tilde k$ for $f_I=0.2$. We see that a large part of excitatory neurons are locked in phase, yielding the quasi-synchronous events appearing also in the inset of Fig. \ref{isiconfr}.
The inhibitory neurons are mainly unlocked and they fire at a higher frequency. In the inset we plot the 
average activity  fields received by  excitatory and inhibitory neurons, $Y_E(t)$  and $Y_I(t)$. Both fields take positive values, while
$Y_I(t) > Y_E(t)$,  as a consequence of the facilitation mechanism that increases the synaptic efficiency during 
fast firing activity. 

For $f_I > 0.7$  (regime C),  $Y_E(t)$ becomes negative and its amplitude reduces significantly,
while both $Y_E(t)$  and $Y_I(t)$ do not  exhibit any oscillating behavior (not shown). This dynamical phase 
is dominated by inhibitory neurons and the natural firing activity of all neurons slows down to an 
irregular behavior, where quasi--synchronous events disappear. 

For  $0.55 <  f_I  < 0.7$ (regime B) partial synchronization with quasi--synchronous events persist, as in regime A.
On the other hand, the microscopic organization of firing events  is  different, as shown in the  middle panel of
Fig.\ref{isiconfr1}, where $f_I = 0.6$. Looking at the inset, one observes that inhibitory neurons receive a relatively 
higher amplitude 
activity field, $Y_I(t)$, with respect to excitatory ones, $Y_E(t)$.  The main distinctive feature with respect to
regime A is that both  $Y_E(t)$ and  $Y_I(t)$ take also negative values, and inhibitory neurons are no more faster than the excitatory ones,
 as one can easily realize looking at the average ISI {\sl vs.} $\tilde k$. In particular, 
an appreciable  subset of  inhibitory neurons lock at the same frequency of {\sl locked} excitatory ones
(see the initial plateau around $\tilde k = 0.4$). As to the {\sl unlocked} inhibitory neurons, they fire at  
lower frequencies with respect to the excitatory ones, that, on their side, are {\sl locked} for large values of $\tilde k$
(compare the upper and middle panels of Fig.\ref{isiconfr1}).

At the edge between regimes A and B, there is a region of {\sl optimal synchronization}
(grey band  in Fig.\ref{sigma}), where neither excitatory nor inhibitory neurons prevail.
The dynamics typical of this region is shown in the lower panel of Fig.\ref{isiconfr1}, where $f_I = 0.5$. 
The average ISI is independent of  $\tilde k$. This means that all neurons fire with a common frequency 
very close to $1/T_0$, i.e. the frequency of the non--interacting system ($g = 0$); only their relative
phases depend on $\tilde k$ . This notwithstanding,
 the microscopic organization of firing events is still quite complex: as shown in the raster plot in the left inset,
there is a majority of inhibitory and excitatory neurons participating the same quasi--synchronous
events, i.e. they are almost in phase.
The right inset shows that $Y_E(t)$ and $Y_I(t)$ exhibit periodic fluctuations of very small amplitude
(i.e. $g \approx 0$), apart 
narrow activity peaks, that correspond to the quasi--synchronous events shown in the left inset.
This is quite an interesting collective dynamical behavior, emerging in a weakly interacting network,
where the complex organization
of the phases of equally periodic excitatory and inhibitory neurons is determined by the 
random structure of the network, i.e. by  the in-degree density 
distributions $P_E(\tilde k)$ and $P_I(\tilde k)$.

\section{Relation between average excitatory and inhibitory activity fields}
\label{quattro}
The HMF approach provides also the possibility of working out an analytic study of the complex network dynamics
described in the previous section. For instance, direct inspection of Fig. \ref{isiconfr1} suggests that the 
average synaptic activity fields defined in Eq.s(\ref{globI}) and (\ref{globE}) exhibit quite a similar behavior in time, 
despite single inhibitory and excitatory neurons receive different synaptic activity fields and emit different spike-trains.
\begin{figure}
\centering
\includegraphics[width=8.1 cm]{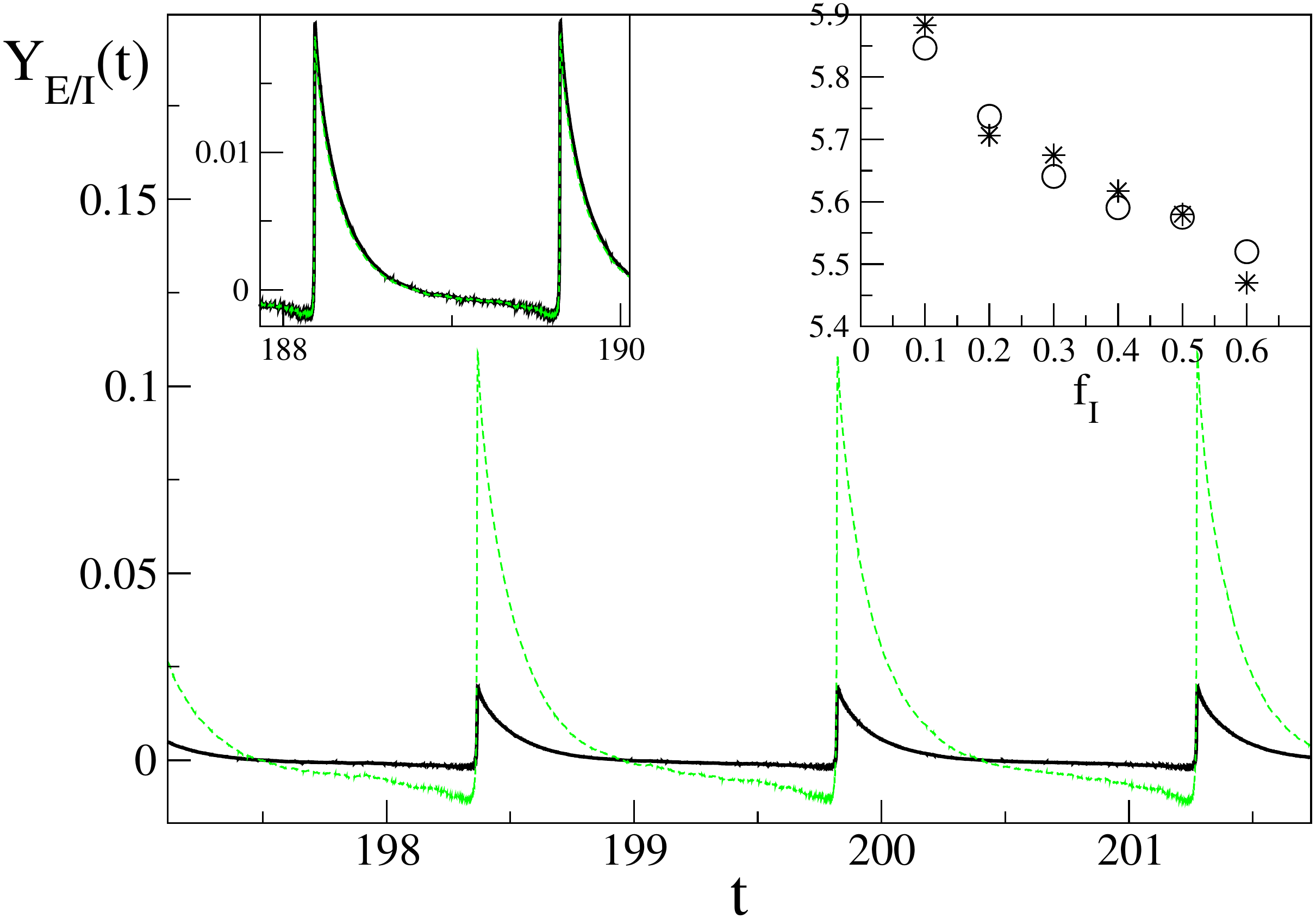}
\caption{Time evolution of  $Y_I(t)$ (green dashed line) and $Y_E(t)$ (black continuous line) for the HMF dynamics with $f_I=0.1$. In the inset on the left, $Y_I(t)$ has been rescaled using the common  period of the global fields and the factor obtained analytically (see text). In the inset on the right we show the comparison between the scale factor deduced from simulations (black circles) and the one obtained analytically (stars). The latter has been calculated by dividing r.h.s. term of Eq. (\ref{prop1}) by r.h.s. term of Eq. (\ref{prop2}), where the period $T$ was obtained from the time evolution of the global fields. The distributions 
$P_{E/I}(\tilde k)$ and the HMF sampling adopted  are the same of Fig. \ref{isiconfr}. } 
\label{scaling}
\end{figure} 
In Fig. \ref{scaling} we focus on the time evolution of $Y_E(t)$ and $Y_I(t)$ for the HMF with $f_I=0.1$ (See Figure caption for details).  
Apart from small fluctuations, the two fields  coincide by a suitable rescaling of their amplitudes (see left inset). 
An analytic estimate of the rescaling factor can be obtained by a heuristic argument. 
Since $Y_E$ and $Y_I$ depend on the fields $y^{(E, *)}_{\tilde k}$ and
$y^{(I, *)}_{\tilde k}$, respectively (see Eq.(\ref{ygrandestar})), their difference 
can be traced back to the different dynamical behavior of  $u_{\tilde k}^{(\dagger,*)} $ (see Eq.(\ref{uk})),
that comes into play at the firing events. Accordingly, in between two spikes,  $Y_I(t)$ and $Y_E(t)$ follow the same dynamics, 
i.e. an exponential decay with the same time constant $\tau_{\mathrm{in}}$. 
Let us consider a neuron that, firing its spike train, generates postsynaptic fields $y^{\dagger,*}_{\tilde k}(t)$
and assume that  it emits spikes at a constant rate, i.e. its synaptic activity field is periodic with the same period $T$ 
of the average activity  field (actually, locked neurons display such a behavior). By imposing the periodicity
properties to  Eq.s (\ref{yk}) and (\ref{uk}) i.e. $y^{\dagger,*}_{\tilde k}(t)=y^{\dagger,*}_{\tilde k}(t+T)$, we can
obtain an explicit expression of their time  dependence. Both fields  exhibit  the same exponential decay,
with time constant $\tau_{\mathrm{in}}$, and their amplitudes  are found to depend on the
different boundary conditions at  firing events for excitatory and inhibitory neurons.  
In formulae we report their maximum values, $\tilde y^{E,*}_{\tilde k,MAX}$ and $\tilde y^{I,*}_{\tilde k,MAX}$, 
achieved immediately after the spike emission:

\begin{align}
\label{prop1}
&\tilde y^{E,*}_{\tilde k,MAX}= \frac{U } {1-e^{-\frac{T}{\tau_{\mathrm{in}}}}\Big(1+U+\frac{U\tau_{\mathrm{r}}^E}{\tau_{\mathrm{r}}^E-\tau_{\mathrm{in}}}\Big)\Big(e^{-\frac{T}{\tau_{\mathrm{r}}^E}}e^{\frac{T}{\tau_{\mathrm{in}}}}-1   \Big)    }\\
\label{prop2}
&\tilde y^{I,*}_{\tilde k,MAX}=  \frac{\tilde u^{I} }{1-e^{-\frac{T}{\tau_{\mathrm{in}}}}\Big(1+\tilde u^{I}+\frac{\tilde u^{I}\tau_{\mathrm{r}}^I}{\tau_{\mathrm{r}}^I-\tau_{\mathrm{in}}}\Big)\Big(e^{-\frac{T}{\tau_{\mathrm{r}}^I}}e^{\frac{T}{\tau_{\mathrm{in}}}}-1   \Big)    }\\
\label{prop3}
&\tilde u^{I}=U_f\frac{e^{-\frac{T}{\tau_f}}}{1-e^{-\frac{T}{\tau_f}}+U_fe^{-\frac{T}{\tau_f}}}.
\end{align}

Notice that, as the two fields decrease exponentially with the same time constant in between two consecutive spikes,
 the field $\tilde y^{E,*}_{\tilde k}(t)$ is equal to $\tilde y^{I,*}_{\tilde k}(t)$ apart from a scaling factor that can be calculated 
at their maximum values, obtained from Eq.s (\ref{prop1})--(\ref{prop2}).
In the right inset of Fig. \ref{scaling} we compare the scaling factor computed numerically with the
analytic prediction obtained from Eq.s(\ref{prop1})--(\ref{prop3}). The agreement is quite good, despite
the simplifying assumptions introduced in the analytic estimate. Let us point out that, in principle, this
should hold for a periodic dynamics. Simulations indicate that it is effective also when the frequencies
of many neurons are not too far from the one of the average activity fields $Y_E(t)$ and $Y_I(t)$.

\section{The Inverse Problem with inhibitory neurons}
\label{cinque}
Global synaptic activity fields in extended regions of the brain can be more accessible to experimental measurements than single--neuron activities. As shown in two previous papers \cite{BCDLV,lungo}, 
in the HMF frame one can recover the degree distribution of a fully excitatory LIF network from the knowledge of its 
global synaptic activity field. In other words, the HMF formulation allows one to solve a global inverse problem.
Here we discuss how to extend such a result to networks made of inhibitory and excitatory LIF neurons.
The method can be extended to networks with different single neuron models.

Let us assume that we have access to the measure of the average activity field received by neurons, i.e.  $Y(t)=f_EY_E(t)+f_IY_I(t)$. We would recover, within a reasonable accuracy,  $P_E(\tilde k)$, $P_I(\tilde k)$ and $f_I$,
i.e. the probability distributions of the equivalent in--degree density $\tilde k$ of excitatory and inhibitory
neurons, as well as their fraction. The procedure goes through the following steps.

\vskip .1 cm
{\bf (i)} As discussed in Sec.\ref{quattro}, $Y_E(t)$ and $Y_I(t)$ can be rescaled by a suitable proportionality constant,
$\gamma$, whose explicit expression depends only on phenomenological parameters of the model
(see Eq.s(\ref{prop1})--(\ref{prop3})). Accordingly, we can write 
\begin{equation}
\label{inv}
Y_I(t)= \gamma Y_E(t) = \frac{\gamma Y(t)}{1 + (\gamma -1) f_I}
\end{equation}
and we consider  $Y_E(t)$ and $Y_I(t)$ as  functions of $Y(t)$ and of  the unknown fraction $f_I$;

\vskip .1 cm
{\bf (ii)} For each value of $f_I$, we integrate the dynamics (\ref{vk1}) and (\ref{vk2}),
that produce the spike trains $S_{\tilde k}^*(t)$ that allow to integrate Eq.s (\ref{yk})--(\ref{uk}).

\vskip .1 cm
{\bf (iii)} We use  $\mathsf{y}^{(\dagger, *)}_{\tilde k}(t,f_I)$ to impose the self-consistency 
condition  (see (\ref{ygrandestar}))
\begin{equation}
\label{selfc}
\tilde Y_{\dagger *}(t)=\int^1_0 P_{*}(\tilde k) \mathsf{y}^{(\dagger, *)}_{\tilde k}(t,f_I) d\tilde k,
\end{equation}
We can write the equations analogous to (\ref{globI}) and (\ref{globE})
\begin{align}
\label{globIt}
& \tilde Y_I=-f_I \tilde Y_{II}(t)+f_E \tilde Y_{IE}~, \\
\label{globEt}
& \tilde Y_E=-f_I \tilde Y_{EI}(t)+f_E \tilde Y_{EE} ~.
\end{align}
and $$\tilde Y(t)=f_E \tilde Y_E(t)+f_I \tilde Y_I(t)$$

\vskip .1 cm
{\bf (iv)} Notice that the effective field $\tilde Y(t)$ depends on the quantities to be recovered, namely 
 $P_E(\tilde k)$, $P_I(\tilde k)$ and $f_I$, and the 
 self-consistency condition (\ref{selfc}) should hold only if $\tilde Y(t) \to Y(t)$.
 In practice, a suitable estimate of the unknown quantities can be obtained by minimizing the variance
 \begin{equation}
\sigma^2=\frac 1 {t_1- t_0} \int_{t_0}^{t_1} ( \tilde Y(t)-Y(t))^2 dt
\end{equation}
where $[t_0,t_1]$ is the measurement time interval of $Y(t)$. The minimization procedure can be achieved by a zero temperature Montecarlo algorithm, as for the purely excitatory case (see \cite{BCDLV}). 

\begin{figure}
\centering
\includegraphics[width=8.1 cm]{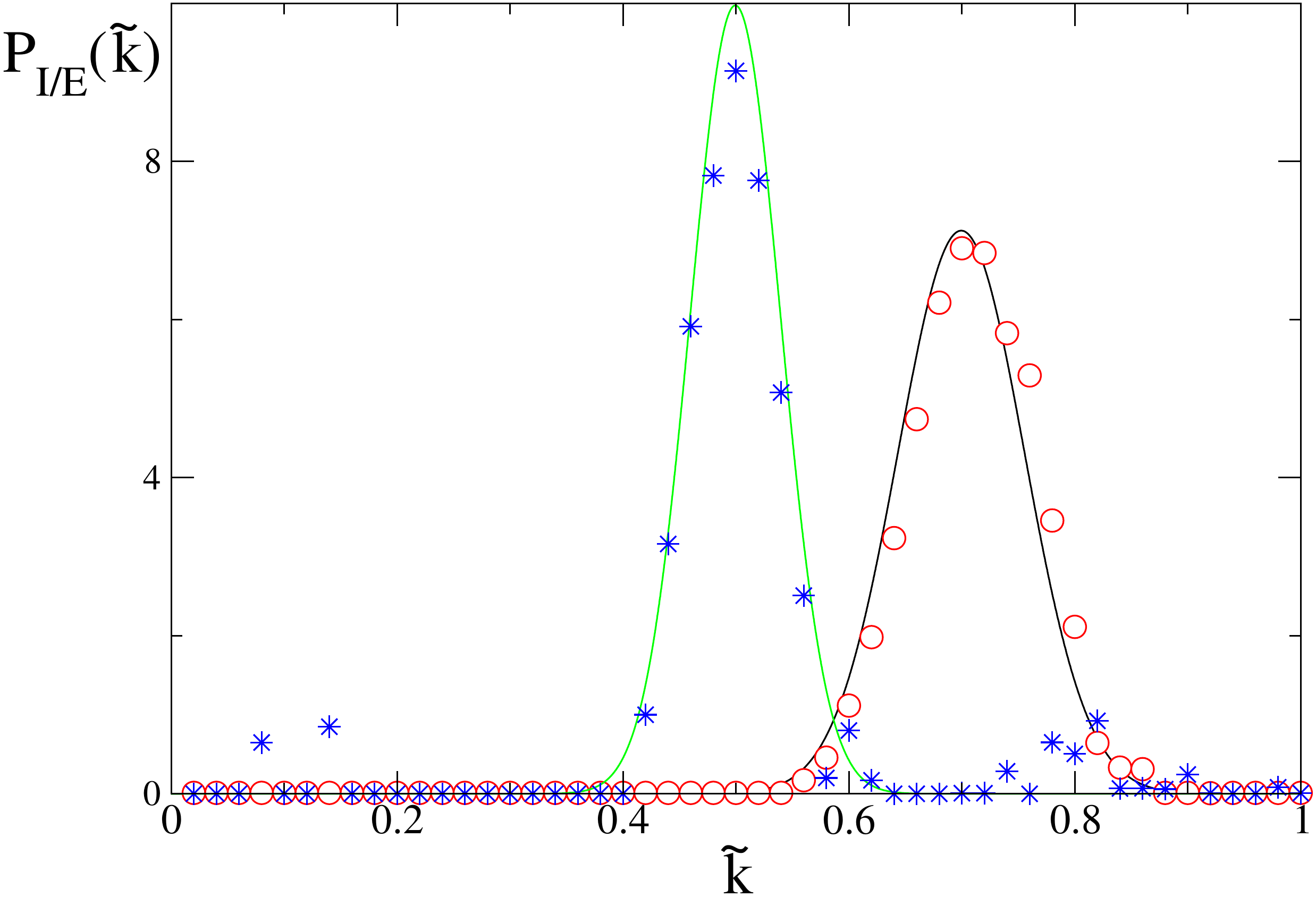}
\caption{ (Color online) Reconstruction of $P_E(\tilde k)$ and $P_I(\tilde k)$ for the network of Fig. \ref{isiconfr}. Continuous curves are the expected distributions while red circles (excitatory neurons) and blue stars (inhibitory neurons) are the reconstructions obtained with the self--consistent inversion equation.} 
\label{distr}
\end{figure} 

The inverse problem procedure {\bf (i)}--{\bf (iv)}
allows to recover quite well the fraction $f_I=0.1$ of inhibitory neurons. In Fig. \ref{distr} we show the reconstruction of $P_E(\tilde k)$, $P_I(\tilde k)$ 
for the dynamics reported in Fig. \ref{isiconfr}. This analysis confirms that, in the case of  Gaussian in--degree density distributions for both excitatory and inhibitory neurons, the average synaptic activity  signal can be efficiently inverted. 

Interestingly, this global inverse procedure can be applied also to the case of 
broad power law distributions, with inhibitory neurons typically displaying  higher connectivities \cite{Boni}. 
As an example, here we report just the case of a network with $f_I = 0.3$ and where 
$P_E(\tilde k)$ and $P_I(\tilde k)$ are power law distributions, scaling as ${\tilde k}^{-\alpha}$ \cite{CHA} . 
In order to avoid too small values of $\tilde k$ we impose a lower cutoff, ${\tilde k_m}^E$ and ${\tilde k_m}^I$, 
over both probability distributions, that are accordingly normalized.

In Fig. \ref{distr_scale} we report the results of the inverse problem procedure. The average
synaptic activity field $Y(t)$ has been computed from the dynamics of a finite network of $N=5000$ neurons. 
In panels A and B we show the reconstruction of $P_E(\tilde k)$ and $P_I(\tilde k)$.  The minimization procedure (step {\bf (iv)} ) 
 provides quite an accurate reconstruction of the fraction  $f_I \approx 0.3$ and 
of the distribution  $P_E(\tilde k)$ over the whole range of definition,
 while $P_I(\tilde k)$ is recovered  just for  $\tilde k \ge {\tilde k_m}^I$ .  This result
 indicates that a more refined algorithm should be employed
 to improve the quality of the inversion. 

We remark that, despite in this case the reconstruction of $P_I(\tilde k)$ is not completely reliable, the presence of a fraction of inhibitory neurons in the inversion procedure is crucial also for the reconstruction of the excitatory distribution.
Let us consider the global field and implement the inverse problem in absence of inhibitory neurons (actually the procedure reported in \cite{BCDLV}). In the lower panel of Fig. \ref{distr_scale},  we can observe that, by  omitting the presence of inhibitory population, the reconstruction of the network structure yields a bad regeneration of $P_E(\tilde k)$ as well (see the inset). Moreover, we compare the reconstruction performance by  plotting the convergence of  $\sigma$ in the Montecarlo minimization. The curve, in presence of only excitatory neurons, converges to higher values of $\sigma$. More precisely, the relative error $\delta =\sigma/\langle Y\rangle$, which is $48\%$ in the case without inhibitory neurons, reduces to $1\%$ when inhibitory neurons are taken into account.  This implies that the matching between $\tilde Y(t)$ and $Y(t)$ is neatly improved by considering the inhibitory population.

\begin{figure}
\centering
\includegraphics[width=8.1 cm]{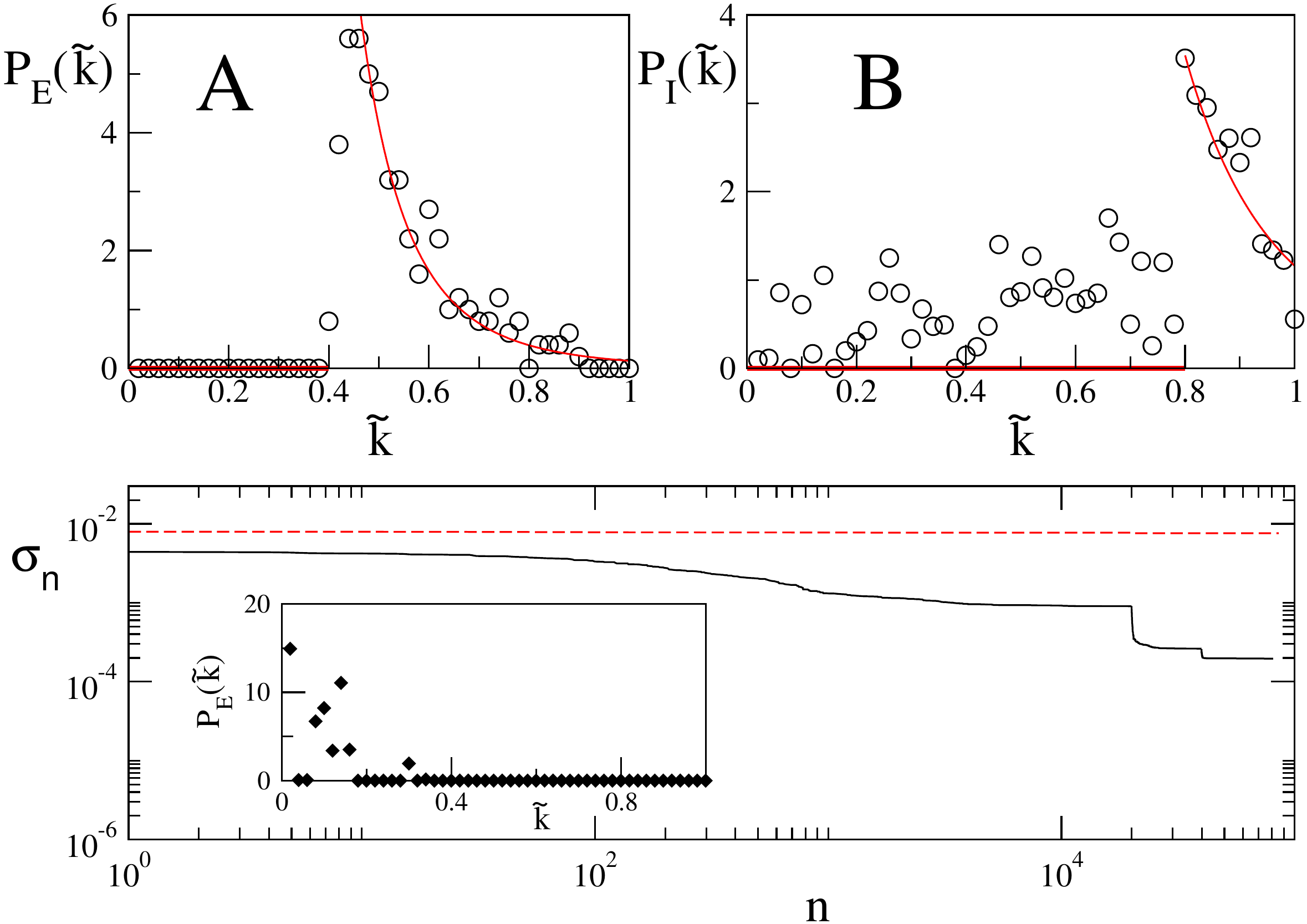}
\caption{ (Color online) Reconstruction with power law connectivity degree distributions. Panels A and B display the reconstructions  of $P_E(\tilde k)$ and $P_I(\tilde k)$ respectively (black circles). The fraction of inhibitory neurons results to be $f_I=0.3$, as wanted. Red continuous line is the power law $\tilde k^{-5}$ with cutoff $\tilde k_m^I=0.8$ and $\tilde k_m^E=0.4$. The lower panel (where $n$ is the MC step) shows the procedure result  when the global field is inverted by considering only excitatory neurons. In particular, we compare  the convergence of the parameter $\sigma$ in the case of panels A and B where the presence of inhibitory neurons is taken into account  (black continuous line) with that in the case where the inverse problem has been performed by considering only excitatory population (dashed red line).  In the inset we show the reconstructed $P_E(\tilde k)$ in this last case and we see that it is quite far from the expected one of panel A.
} 
\label{distr_scale}
\end{figure}


\section{Conclusions and perspectives}
\label{sei}

We have studied the dynamics of random uncorrelated dense networks  of LIF neurons with inhibitory and
excitatory components  by the Heterogeneous Mean Field approximation. This method proves extremely effective in reproducing the complex emerging dynamical phases of the system and provides significative advantages, both in numerical 
simulations and in the analytic approach to the inverse problem, that can be formulated in terms of average properties.

The model presents a very rich dynamical phase diagram where inhibitory and excitatory components feature different complex evolutions. Such a complexity does not restrain the HMF approach from offering an interesting first insight on the global dynamics, that  can be grasped directly from a simplified version of  the HMF equations. Precisely, if the dynamics of the two types of synapses would be the same, the presence of a certain fraction  $f_I$ of inhibitory neurons could be described by introducing an effective in--degree  density distribution. 

Indeed, since in this simple case the same activity fields are transmitted to inhibitory or excitatory  neurons, 
we  just deal with a single set of  evolution equations, where the dependence on the inhibitory
or excitatory nature of presynaptic and postsynaptic neurons can be omitted. In fact, any neuron with degree $\tilde k$  receives a field $g\tilde k Y$, where 
$$Y = \int_0^1\Big[f_EP_E(\tilde k)-f_IP_I(\tilde k)\Big]y_{\tilde k}(t)d\tilde k \quad.$$


If the term in square brackets has a definite sign, the network is equivalent to a completely inhibitory or excitatory (depending on the sign) network with an effective probability distribution $F(\tilde k)=  \Big|f_EP_E(\tilde k)-f_IP_I(\tilde k)\Big|$. In particular, if $P_E=P_I$  the introduction of a fraction of inhibitory neurons $f_I$ is equivalent to an {\it effective dilution} in the original network, obtained through $2f_I$ cuts of the links. If the term in square brackets has no definite sign, $F(\tilde k)$ is not a probability distribution, so that the real dynamics does not correspond to an equivalent excitatory or inhibitory network. 

In the more complex case considered in this paper, the inhibitory dynamics is characterized by a facilitation effect and the fields received by inhibitory neurons turns out to be larger than the excitatory field. This difference can be approximately estimated by an analytic argument, that allows to implement an approximated inverse problem for the distributions $P_*(\tilde k)$ even in the presence of inhibition. 

The global inverse problem proves very effective in reproducing the in--degree density probability distributions of the 
two  populations, together with the fraction of inhibitory neurons, for meaningful distributions ranging 
from Gaussians to power laws. On a technical ground,  the inversion could be improved by adopting more
refined minimization procedures, but it is significant that, even the simple zero--temperature Monte Carlo method adopted here, 
it is sufficient to provide the wanted outcome. 

This result paves the way  to an analysis of  experimental data of the average synaptic  activity fields from broad regions in the brain. 
 Moreover, the overall approach can be extended to models of correlated dense networks and with single neuron dynamics different from LIF.

\begin{acknowledgments}
The authors would like to warmly thank L. De Arcangelis for useful suggestions and discussions
and for the constant stimulus to  converge, as soon as possible, to the publication
of the results contained in this manuscript.

\end{acknowledgments}

\end{document}